\documentclass[10pt,english]{IEEEtran}
\usepackage[LGR,T1]{fontenc}
\usepackage[latin9]{inputenc}
\usepackage[a4paper]{geometry}
\usepackage{array}
\usepackage{float}
\usepackage{booktabs}
\usepackage{textcomp}
\usepackage{mathrsfs}
\usepackage{multirow}
\usepackage{amsmath}
\usepackage{amssymb}
\usepackage{graphicx}

\makeatletter

\ProvideTextCommand{\~}{LGR}[1]{\char126#1}

\providecommand{\tabularnewline}{\\}
\floatstyle{ruled}
\newfloat{algorithm}{tbp}{loa}
\providecommand{\algorithmname}{Algorithm}
\floatname{algorithm}{\protect\algorithmname}

\usepackage{flushend}

\makeatother

\usepackage{babel}
\begin{document}
\title{\Large \bf MAED: Mathematical Activation Error Detection for Mitigating Physical Fault Attacks in DNN Inference}
\author{Kasra Ahmadi, Saeed Aghapour, Mehran Mozaffari Kermani, \textit{Senior
Member}, \textit{IEEE, }and Reza Azarderakhsh\textit{, Member, IEEE}}
\IEEEspecialpapernotice{\thanks{K. Ahmadi, S. Aghapour, and M. Mozaffari-Kermani are with the Department
of Computer Science and Engineering, University of South Florida,
Tampa, FL 33620, USA. e-mails: \{ahmadi1, aghapour, mehran2\}@usf.edu.\protect
\protect \\
 R. Azarderakhsh is with the Department of Computer and Electrical
Engineering and Computer Science, Florida Atlantic University, Boca
Raton, FL 33431, USA. e-mail: razarderakhsh@fau.edu.}}
\maketitle
    
\maketitle

%-------------------------------------------------------------------------------
\begin{abstract}
The inference phase of deep neural networks (DNNs) in embedded systems is increasingly vulnerable to fault attacks and failures, which can result in incorrect predictions.
These vulnerabilities can potentially lead to catastrophic consequences, making the development of effective mitigation techniques essential.
In this paper, we introduce \textit{MAED} (Mathematical Activation Error Detection), an algorithm-level error detection framework that exploits mathematical identities to continuously validate the correctness of non-linear activation function computations at runtime.
To the best of our knowledge, this work is the first to integrate algorithm-level error detection techniques to defend against both malicious fault injection attacks and naturally occurring faults in critical DNN components in embedded systems.
The evaluation is conducted on three widely adopted activation functions, namely ReLu, sigmoid, and tanh which serve as fundamental building blocks for introducing non-linearity in DNNs and can lead to mispredictions when subjected to natural faults or fault attacks.
We assessed the proposed error detection scheme via fault model simulation, achieving close to 100\% error detection while mitigating existing fault attacks on DNN inference.
Additionally, the overhead introduced by integrating the proposed scheme with the baseline implementation (i.e., without error detection) is validated through implementations on an AMD/Xilinx Artix-7 FPGA and an ATmega328P microcontroller, as well as through integration with TensorFlow. On the microcontroller, the proposed error detection incurs less than 1\% clock cycle overhead, while on the FPGA it requires nearly zero additional area, at the cost of approximately a 20\% increase in latency for sigmoid and tanh. The proposed defense mechanism, when combined with weight protection schemes, provides a robust solution for mitigating fault attacks in DNNs utilized in embedded systems.

\end{abstract}

\section{Introduction}
The integration of machine learning (ML) into the foundational fabric of modern society necessitates a rigorous focus on safety as these systems transition from digital environments to physical world actuators. ML models now serve as the cognitive backbone for critical domains such as autonomous transportation~\cite{sato2021dirty}, smart city infrastructure~\cite{smartcity}, diagnostic healthcare~\cite{panahi2025deep}, and industrial automation~\cite{industrysurvey}.

In these mission critical settings, traditional metrics of predictive accuracy are insufficient; the emphasis must instead shift toward functional safety and resilience.
As ML systems proliferate across the global landscape, their operational integrity is increasingly threatened by a broad spectrum of fault sources, ranging from stochastic hardware induced errors, such as bit flips, to sophisticated adversarial perturbations.
In these critical scenarios, a single point of failure poses an unacceptable risk to critical infrastructure and can precipitate catastrophic consequences, including economic instability or the loss of human life.
As a result, designing robust error detection and fault tolerance mechanisms that enable ML systems to fail safely under adverse conditions has emerged as a major challenge in modern research.
Traditional attacks on DNNs focus on compromising either model integrity during the training phase or data integrity during the inference phase. In contrast, they overlook the integrity of the hardware on which DNN models are deployed and executed. 

In ML systems, transient faults can have serious consequences during the training process~\cite{google_fault_training}. 
Hardware failures are an increasingly significant issue in data centers, as reflected by the rising number of incidents recently reported by companies such as Google, Meta, and others \cite{cloud_1,cloud_2,cloud_3}. 
As DNNs are increasingly deployed on embedded systems for training and inference, these systems become more vulnerable to hardware failures and fault attacks~\cite{edge_1,edge_2,edge_3,edge_4}.

Recent studies have introduced a new class of fault injection attacks~\cite{proflip,knock,terminal} that violate model integrity by targeting the underlying hardware. These attacks manipulate the deployed model in a manner analogous to training phase attacks, but crucially, they occur after deployment. Several works \cite{attack_1,attack_2} examine how Fault Injection Attacks (FIAs) affect embedded systems running DNNs, where attackers attempt to manipulate the network's parameters to force a specific input to be misclassified into a targeted adversarial class.
Activation functions are widely utilized across different DNN architectures and Transformers~\cite{vaswani2017attention}, playing a crucial role in introducing non-linear behavior into the models~\cite{goodfellow2016deep}.
The vulnerabilities of common activation functions, including softmax, ReLu, sigmoid, and tanh, to FIAs are the primary focus of study~\cite{attack_1}.
There is a pressing need to develop efficient and lightweight mechanisms to protect DNN inference from natural faults and FIAs, particularly in Edge AI devices, which are highly vulnerable to such threats.
\section{Preliminaries}
\subsection{Hardware Faults}
Hardware faults pose a major challenge for both conventional and ML systems.
They arise when physical components, such as processors, memory units, storage devices, or interconnects, malfunction or exhibit abnormal behavior.
Such faults can lead to incorrect computations, data corruption, exposure of cryptographic keys, and ML model mispredictions, ultimately undermining the integrity and reliability of the system's computations.
Faults can generally be divided into two main categories. 

The first type is transient faults, which are temporary and do not recur. 
These are usually triggered by external factors such as electromagnetic interference or power fluctuations. 
A typical example is a bit-flip, where a single bit in memory or a register unexpectedly changes its value.
In ML systems, transient faults can greatly impact both the training and inference phases.
For example, a single bit-flip in a neural network's weight matrix may cause the model to learn incorrect relationships or patterns, resulting in reduced accuracy \cite{transient_bit_flip_NN}.
Similarly, faults within the data pipeline, such as corrupted training samples or mislabeled data, can introduce noise and degrade the overall quality of the trained model.
The second type is permanent faults, which are irreversible and persist over time. 
They usually result from physical damage or the gradual degradation of hardware components.
Common examples include stuck-at faults, where a bit or signal is permanently fixed at a specific value (e.g., 0 or 1).

\subsection{Fault Injection Attacks (FIAs)}
Historically, FIAs have been a cornerstone of side-channel analysis in cryptographic implementations, primarily aimed at extracting secret keys through techniques such as Differential Fault Analysis (DFA)~\cite{ravi2024side}. However, recent research has demonstrated a critical shift in the threat landscape, moving from the retrieval of static cryptographic secrets to the intentional manipulation of DNN inference behavior~\cite{attack_1, attack_2}. 

FIAs represent a potent class of physical attacks that deliberately induce transient or permanent hardware faults during system execution. By perturbing intermediate computations, attackers can precisely alter internal data paths, leading to controlled deviations in results while bypassing conventional software level protections. Well established methodologies, including laser based injection~\cite{laser_inject} and RowHammer~\cite{RowHammer} based attacks, enable highly localized modifications at the logic or memory level.

When applied to DNNs, these attacks compromise inference integrity by altering neuron activations or weight parameters stored in memory. Table~\ref{tab:FIA} summarizes the existing FIAs targeting DNNs.
Among these, the Bit-Flip Attack (BFA)~\cite{bit-flit-attack} is regarded as one of the most severe threats.
BFA~\cite{bit-flit-attack} and DeepHammer~\cite{yao2020deephammer} both exploit Rowhammer induced faults at the model weight level and assume full white-box access, but they differ in how they identify and leverage vulnerable bits. The BFA relies on a progressive search strategy that individually flips weight bits and empirically measures their impact on inference accuracy, ultimately selecting bits that cause the largest degradation when corrupted. In contrast, DeepHammer adopts a more structured, analytical approach by modeling fault propagation across layers, identifying chains of flipped bits whose effects amplify as they traverse the network. 
While both attacks require the simultaneous injection of many bits (typically over a dozen) to achieve meaningful accuracy loss, DeepHammer improves attack efficiency by exploiting inter layer dependencies rather than treating bit-flips independently. Nevertheless, both approaches remain model dependent, require precise knowledge of the network architecture and parameters, and are sensitive to model updates, which fundamentally limits their scalability and generality.

FrameFlip~\cite{frameflip} introduces a new attack that exhausts DNN inference by injecting runtime faults into the executing code. A key advantage of FrameFlip is its model agnostic nature: it succeeds with a single bit-flip and does not rely on the specific DNN architecture being deployed. This sets it apart from prior attacks, which typically require the simultaneous injection of dozens of deterministic faults. FrameFlip targets universal code at the library level, exploiting code paths that are inevitably invoked by major machine learning frameworks such as PyTorch~\cite{pytorch}, TensorFlow~\cite{tensorflow}, and Caffe~\cite{caffe}. The attack leverages DRAM Rowhammer to enable end-to-end fault injection directly.
%~\cite{caffe}
\begin{table*}[t] % Adding the * makes it span two columns
\centering
\caption{Comparison of FIAs targeting DNNs}\vspace{0.5mm}
\label{tab:FIA}
\resizebox{\textwidth}{!}{% % Changed columnwidth to textwidth
\begin{tabular}{ccccccccc}
\toprule
 &
  \begin{tabular}[c]{@{}c@{}}Attack \\ Level\end{tabular} &
  \begin{tabular}[c]{@{}c@{}}Attack \\ Vector\end{tabular} &
  \begin{tabular}[c]{@{}c@{}}Required\\ Faulty Bits\end{tabular} &
  \begin{tabular}[c]{@{}c@{}}Model \\ Knowledge\end{tabular} &
  \begin{tabular}[c]{@{}c@{}}Target \\ Environment\end{tabular} &
  \begin{tabular}[c]{@{}c@{}}Physical \\ Access\end{tabular} &
  \begin{tabular}[c]{@{}c@{}}Protected by\\  \textit{MAED}\end{tabular} &
  Transferability \\ \hline
Frameflip~\cite{frameflip} &
  \begin{tabular}[c]{@{}c@{}}Code Library\\ (BLAS)\end{tabular} &
  \begin{tabular}[c]{@{}c@{}}Rowhammer\\ (Software)\end{tabular} &
  1 &
  Black-box &
  Cloud/MLaaS &
  No &
  \begin{tabular}[c]{@{}c@{}}Potentially\\ (if it affects activation output)\end{tabular} &
  High \\ \hline
BitFlip~\cite{bit-flit-attack} &
  Model Weights &
  \begin{tabular}[c]{@{}c@{}}Rowhammer\\ (Software)\end{tabular} &
  11-20 &
  White-box &
  Cloud/MLaaS &
  No &
  No &
  Low \\ \hline
DeepHammer~\cite{yao2020deephammer} &
  Model Weights &
  \begin{tabular}[c]{@{}c@{}}Rowhammer\\ (Software)\end{tabular} &
  11-23 &
  White-box &
  Cloud/MLaaS &
  No &
  No &
  Low \\ \hline
DeepLaser~\cite{attack_1} &
  Activation Functions &
  \begin{tabular}[c]{@{}c@{}}Laser Injection\\ (Hardware)\end{tabular} &
  5-15 &
  Gray-box &
  Embedded Systems/IoT &
  Yes &
  Yes &
  Medium
  \\ \bottomrule
\end{tabular}
}
\end{table*}

\subsection{Defenses Against Fault Attacks}
To reduce hardware faults and mitigate FIAs, ML systems need to integrate fault detection mechanisms at the hardware or software levels.
Hardware redundancy \cite{redundancy_survey} is an approach which relies on replicating critical components and comparing their outputs to detect faults.
Voting schemes such as double modular redundancy (DMR) and triple modular redundancy (TMR) deploy multiple instances of the same component and evaluate their outputs to identify and suppress erroneous behavior.
In DMR and TMR architectures, two or three identical hardware units execute the same operation concurrently.
Their outputs are then provided to a voting logic that compares the results and determines the final output based on agreement or majority.
When a fault causes one instance to produce an incorrect result, the voting mechanism masks the error and preserves correct system operation.
TMR is widely adopted in aerospace and aviation applications, where stringent reliability requirements are paramount.
By contrast, Tesla's self driving computers adopt a DMR based design to enhance the safety and dependability of critical functionalities, including perception, decision making, and vehicle control.

Algorithm-level error detection serves as a software based alternative for fault detection; this method identifies execution errors by leveraging the algorithm's intrinsic mathematical or structural properties rather than relying on dedicated hardware overhead.
Rather than duplicating hardware units, it integrates validation steps within the algorithm itself, such as verifying intermediate computations, preserving invariants, or performing redundant calculations based on algorithmic properties.
Algorithm-level error detection is especially beneficial in both high performance and resource constrained environments, as it improves system reliability with minimal overhead and is commonly used to safeguard cryptographic~\cite{NTT} and numerical computations against transient or injected faults.
In contrast, hardware based techniques require extensive control over the system and low level hardware access, making them more difficult to implement. 
Conversely, software based solutions offer greater flexibility and can be applied to hardware agnostic scenarios, making them more adaptable across various platforms.

Aegis~\cite{aegis} is designed to defend against targeted BFA, in which an adversary flips carefully chosen weight bits to induce a specific misclassification. Rather than always processing inputs through the full network, Aegis inserts internal classifiers (ICs) at multiple hidden layers and randomly chooses the exit point used for inference. 
% This unpredictability prevents the attacker from knowing which layer will generate the final output, thereby obscuring the critical weights they aim to manipulate. 

DeepDyve~\cite{deepdyve} uses a dynamic verification approach in which a lightweight auxiliary checker model runs in parallel with the primary DNN. The checker is trained to produce an approximate solution to the same task, and a fault or attack is flagged when the primary model's output diverges noticeably from the checker's prediction. 

Weight Reconstruction~\cite{weightreconstruction} counters malicious BFA by mathematically rebuilding weights just before inference to neutralize or diffuse their impact. The approach first groups weights into small grains and computes the average of each grain, which spreads the effect of a flipped bit across multiple weights and reduces localized deviations. The computed mean is subsequently quantized to the closest predefined level, causing minor perturbations to be mapped back to their original values. Finally, all weights within each grain are clipped to a narrow range around this quantized mean, preventing any single corrupted weight from exerting a disproportionate influence on the layer's output.

For DNNs executed on floating point hardware, only a limited portion of the numerical dynamic range supported by the data type is typically utilized. Consequently, faults that cause intermediate activations to grow excessively large are likely to result in erroneous behavior. Leveraging this insight, Threshold checking~\cite{rangedefence} used a lightweight anomaly detection mechanism based on simple thresholding. Specifically, during a pre-deployment profiling phase, the minimum and maximum output values $(X_{\min}, X_{\max})$ are recorded for each network layer. During runtime operation, the layer outputs are continuously monitored and a fault is flagged if any value exceeds the expanded bounds $(1.1 \times X_{\min},\, 1.1 \times X_{\max})$. 
% While this approach incurs negligible hardware overhead, its fault detection effectiveness is limited, particularly against modern attack techniques targeting quantized DNN implementations.

\subsection{Activation Functions}
Activation functions are essential elements in neural network architecture.
Their primary purpose is to introduce non-linearity into the network's computations.
Without this non-linear step applied after each node's weighted sum of inputs, even a deep network with many layers would effectively collapse into a single linear transformation.
Since combining linear functions results in another linear function, the network would be unable to capture complex patterns that cannot be separated by a simple hyperplane.
By introducing non-linearity, activation functions allow neural networks to approximate a wide range of functions, enabling them to act as universal function approximators.
Previous studies~\cite{attack_1,attack_2} show that faults in activation functions are more likely to cause misclassifications than errors in linear layers because of the way neural networks process information.
Small errors in a linear layer tend to propagate proportionally and can be partially absorbed by subsequent layers.
In contrast, activation functions introduce non-linearity, and a fault in an activation can dramatically distort the signal that feeds into the next layer. 
Many activation functions, such as ReLu, sigmoid, or tanh, have regions where the output changes sharply or saturates.
In edge inference and embedded systems, such as microcontrollers, sigmoid and tanh functions are often approximated using techniques like Taylor series expansions~\cite{Taylor}.
This approach avoids expensive exponential computations and allows for more efficient computation on resource constraint devices.

In practice, activation functions are frequently implemented using approximation techniques to satisfy the stringent power and latency requirements of resource constrained devices. A prevalent method for computing the exponential component within floating point systems is the Maclaurin series expansion, which approximates transcendental functions through a finite sum of polynomial terms. For example, to evaluate the sigmoid function, the value $e^{-x}$ is first approximated, and the output is then computed as $y=\frac{1}{1+e^{-x}}$. For the tanh function, additional approximation techniques can be used depending on the application, including infinite products, continued fractions, and Chebyshev or Padé approximations.

\section{DeepLaser and Proposed Countermeasure}
\subsection{DeepLaser FIA}
DeepLaser~\cite{attack_1} is the first practical study into physical FIA on DNNs, demonstrating laser based attacks on embedded systems and closing the gap between theoretically identified DNN vulnerabilities and their feasibility in real world systems. While earlier studies examined fault injection either through software based simulations of neural networks or in isolation on cryptographic hardware, DeepLaser uniquely unifies these lines of work by applying laser based fault injection to microcontroller based DNN implementations, demonstrating that physical attack techniques long established in cryptography pose a direct threat to DNN inference as well.
Instead of manipulating data such as neurons' weights, DeepLaser targets the control-flow of activation functions by inducing instruction skips via near-infrared diode pulse laser injections. The attack realizes standard activation functions (ReLu, sigmoid, tanh, and softmax) on microcontrollers and injects faults by accurately timing laser pulses to skip or alter specific assembly instructions during their execution.

In the case of ReLu, instruction skips are used to force the output to zero irrespective of the input value. For sigmoid and tanh, the attack disrupts the negation step in the exponent calculation, leading to fundamentally incorrect outputs. This approach relies on detailed manual inspection of the compiled assembly code to identify vulnerable instructions, along with precise timing calibration to fault specific operations.
DeepLaser is especially critical for embedded AI and IoT platforms deployed in physically exposed, critical environments where conventional software defenses offer no protection against physical adversaries. Unlike cloud based deployments that benefit from isolation and shielding, edge devices typically lack robust physical security, making the DeepLaser threat model both realistic and alarming. These factors emphasize the critical need for effective fault detection and mitigation strategies, which motivates our choice of this attack model, introduces \textit{MAED}, and highlights the urgent necessity for defenses to safeguard DNN inference against physical FIA.
\subsubsection{DeepLaser Threat Model}
% The study\cite{usenix_bitflip} demonstrates that in most models, there exists at least one parameter whose single bit-flip in its binary representation can result in an accuracy degradation exceeding 90\%.
% %
% For larger models, simple heuristic analysis suggests that around 50\% of parameters could individually trigger an accuracy drop of more than 10\% when exposed to a single-bit fault.
% %
% The study further analyzes this vulnerability by examining the influence of multiple factors, including bit position, flip direction, parameter sign, layer width, activation function, training method, and model architecture.
% %
% To highlight the real-world risk, it employs software-based fault injection through the Rowhammer technique, demonstrating the practical feasibility of bit-flip attacks.\\
%
The DeepLaser~\cite{attack_1} threat model considers a determined adversary who seeks to undermine the correctness of a system's output by injecting faults at the hardware level.
The attacker's primary objective is to cause misclassification during the inference stage.
To this end, the adversary follows a random fault model, aiming to prevent the network from producing the correct label without necessarily enforcing a specific target class.
This setting corresponds to a gray-box adversarial model in which the attacker has partial system knowledge.
In particular, the attacker lacks access to the model's internal parameters, including the network architecture, weight values, and the number of neurons in hidden layers. However, the adversary can exploit side-channel information, such as power consumption or electromagnetic emissions, to infer which activation functions are being executed. However, distinguishing between similar activation functions in term of computation (e.g., sigmoid vs. tanh) may be challenging.
Additionally, by leveraging trigger signals or timing related side-channel leakage, the attacker can accurately identify when a particular layer or operation is processed, enabling precise synchronization of the fault injection.

The attacker is assumed to have both the capability and expertise to physically interfere with the embedded system hardware, requiring direct access to the device under test to perform fault injections.
In advanced attack scenarios, the adversary employs high precision techniques such as diode-pulse laser fault injection~\cite{laser_attack} to target the silicon die with fine spatial and temporal accuracy, enabling the modification or skipping of specific assembly instructions during activation function execution.
This includes instruction-level manipulation, such as bypassing control-flow operations to force a ReLu function to always output zero, thereby effectively disabling a neuron irrespective of its input.
Although high-end laser equipment is used in the study, the threat model also encompasses more accessible, low-cost fault injection methods, including voltage glitching, clock glitching, and electromagnetic attacks, that can achieve similar effects.
To maximize attack effectiveness, the adversary focuses on activation functions in hidden layers closest to the output and is assumed to be capable of injecting multiple faults within a single layer to increase the likelihood of misclassification, particularly in networks with a large number of neurons. The adversary is assumed to have prior knowledge of the activation function types.
\subsection{Proposed Countermeasure}
Since DeepLaser targets activation functions, our proposed error detection scheme is specifically designed for them.
\textit{MAED} utilizes algorithm-level error detection to verify outputs with minimal computational overhead for sigmoid and tanh. 
As discussed in section 2.4, to calculate tanh and sigmoid functions we suppose the baseline (without error detection) and proposed scheme utilizes Taylor series based implementation.
We first use a Taylor expansion to approximate $e^x$ and then use that approximation to evaluate the tanh and sigmoid functions.
Due to the lower computational complexity of the ReLu function relative to tanh and sigmoid, we selected recomputation based with encoding approach~\cite{recomputinoperands} to detect faults during its execution.
The error detection workflow for sigmoid and tanh functions consists of the following three stages:
\begin{enumerate}
    \item \textbf{Baseline Implementation:} The function $y = f(x)$ is computed using the standard high-precision implementation.
    \item \textbf{Inverse Transformation:} We apply a transformation $h(y)$ to the output. This transformation is chosen such that $h(y)$ simplifies to a specific mathematical term, $g(x)$, which was originally part of the function's definition (e.g., an exponential term).
    \item \textbf{Independent Comparison:} We compute an approximation of $g(x)$, denoted as $\hat{g}(x)$, using a computationally efficient method such as a Maclaurin series expansion. An error is detected if the residual between the transformed output and the independent approximation exceeds a predefined threshold $\epsilon$:
    \[
    |h(y) - \hat{g}(x)| > \epsilon
    \]
\end{enumerate}
By caching terms of the Maclaurin series, this method provides a high degree of error coverage with minimal hardware overhead.
\subsubsection{Proposed Countermeasure for Sigmoid}
By exploiting the mathematical definition of the sigmoid function,
\vspace{1mm}
\[
y=\frac{1}{1+e^{-x}}
\]
the correctness of its output can be validated through the following relation:
\[
h_1(y)=\frac{y}{1-y}=e^x
\]
To avoid complete recomputation of the exponential term, \(e^x\) can be efficiently approximated using the Maclaurin series (a Taylor expansion centered $a=0$) at \(m\) terms,
\[
e^{-x} \approx \sum_{k=0}^{m} (-1)^k \frac{x^k}{k!}
\]
and similarly, 
\[
\hat{g}_1(x) = e^{x} \approx \sum_{k=0}^{m}\frac{x^k}{k!}
\]
It is worth noting that by caching or reusing the computed terms of each Maclaurin series expansion, the overall computational cost can be further reduced.
\subsubsection{Proposed Countermeasure for Tanh}
We adopt a similar approach to that used for the sigmoid function. The tanh function can be expressed as:
\[
y = \frac{e^{2x}-1}{e^{2x}+1}.
\]
To verify the correctness of its output, we define,
\[
h_2(y) = \frac{\alpha - 1}{\alpha + 1}, \quad \text{where} \quad \alpha = \frac{1 - y}{1 + y}
\]
This allows validation of the output $y$ by ensuring the following identity holds using the calculated $\alpha$:
\[
\hat{g}_2(x) =\frac{e^{-2x}-1}{e^{-2x}+1}
\]
% If we use Maclaurin series (a Taylor expansion centered at $a=0$) with \(m\) terms, then:
% \[
% e^{-x} \approx \sum_{k=0}^{m} (-1)^k \frac{x^k}{k!}
% \]
% Noting the identity \(y^{-1}-1=e^{-x}\), we obtain:
% \[
% y^{-1}  + x  \approx 2+\sum_{k=2}^{m} (-1)^k \frac{x^k}{k!}
% \]
% This relation indicates that the sum of the inverse sigmoid output and the input is directly related to the partial terms of the Maclaurin series.
% %
% By storing the individual Maclaurin terms during computation, one can reconstruct this series to validate the computed sigmoid output.
% %
% This approach allows the detection of potential errors in the summation process of the sigmoid function, assuming that the computation of each individual term, including the power and division by precomputed factorials, is error-free.
\subsubsection{Proposed Countermeasure for ReLu}
Since the ReLu function is computationally lightweight compared to the tanh and sigmoid functions, we chose to apply recomputation-based methods to detect potential faults during its operation.
The ReLu function is defined as:
\[
\text{ReLu}(x) = max(0,x)
\]
The fault attack described in \cite{attack_1} targets the ReLu function by forcing its output to remain constantly zero.
The correctness of the ReLu function can be verified using a recomputation technique with negated operands, where \(h_3(x) = 0\) denotes the absence of an error and \(h_3(x) = 1\) indicates that an error has occurred.
\[
h_3(x) =
\left\{
\begin{array}{ll}
0, & \text{if } \text{ReLu}(x) + \text{ReLu}(-x) \neq 0 \text{ and } x \neq 0 \\[6pt]
0, & \text{if } \text{ReLu}(x) + \text{ReLu}(-x) = 0 \text{ and } x = 0 \\[6pt]
1, & \text{otherwise (error detected)}
\end{array}
\right.
\]
The effectiveness of our proposed error detection scheme against DeepLaser is described in Table~\ref{tab:effectiveness_fault_attack}. Our simulation and implementation are open-source for public verification and testing\footnote{https://github.com/KasraAhmadi/Error-Detection-DNN}.
\begin{table}[]
\centering
\caption{Effectiveness of \textit{MAED} against DeepLaser attack~\cite{attack_1}}
\label{tab:effectiveness_fault_attack}
\begin{tabular}{@{}ccc@{}}
\toprule
        & Effect of attack                                                           & \begin{tabular}[c]{@{}c@{}}Detected through\end{tabular} \\ \midrule
sigmoid & \begin{tabular}[c]{@{}c@{}}Skipping the negation\\  in the exponent\end{tabular} & \(|h_1(y) - e^x| > \epsilon    \)                                                                        \\ \midrule
tanh    & \begin{tabular}[c]{@{}c@{}}Skipping the negation\\  in the exponent\end{tabular} & \(|h_2(y)- \frac{e^{-2x}-1}{e^{-2x}+1}| > \epsilon\)                                                  \\ \midrule
ReLu    & Always maps to 0                                                                 & \(h_3(x)=1\)                                                                                       \\ \bottomrule
\end{tabular}
\end{table}
\section{Error Coverage and Simulation Results}
In this section, we evaluate the error-coverage ratio of our design under several realistic fault-injection scenarios. Specifically, we consider two types of fault injection and five different injection models, each tested under varying numbers of injected faults. After describing simulation settings and each of these scenarios in detail, we present the detection ratio achieved by the proposed architecture.

\subsection{Methodology and Simulation Details}
In our simulation implementation, we use the IEEE double-precision floating-point format, where each number is represented using 64 bits: the first bit is the sign bit, the next 11 bits are the exponent, and the remaining 52 bits form the significant. 
Furthermore, we assume that faults are injected during the computation of the activation functions, specifically in the register that holds the intermediate value $\frac{x^{k}}{k!}$. We consider the input range of $[-3, 3]$. In addition, each fault-injection model is simulated over 1000 runs, with every run using a new random input and a new set of injected faults at new locations.

\subsubsection{Round-off Error and Taylor Terms}
\begin{figure}[tbp]
    \centering
    \includegraphics[width=0.4\textwidth, trim=10cm 2.5cm 7cm 4cm, clip]{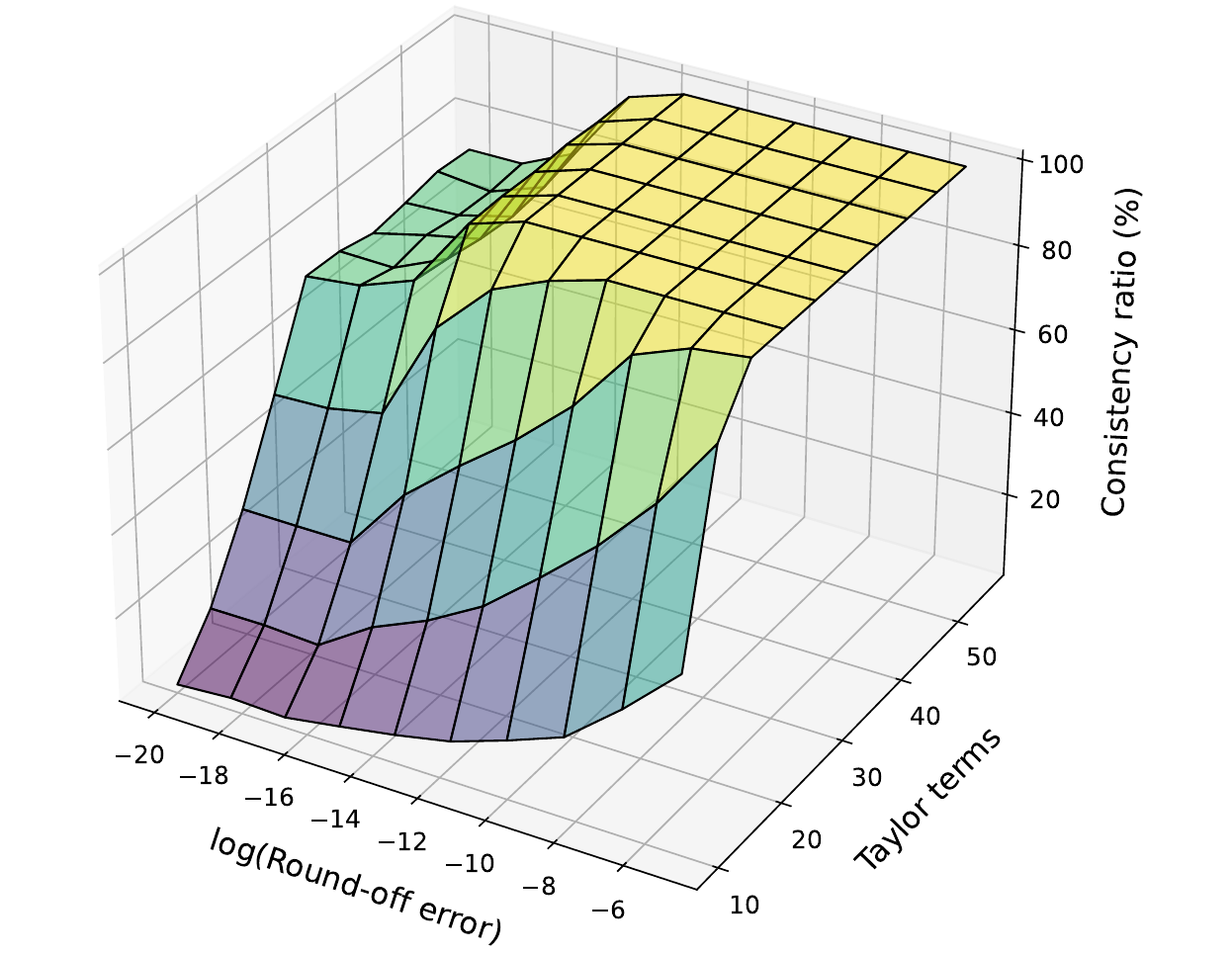}
    \caption{Effect of round-off error and number of Taylor terms to the accuracy of the proposed architecture for exponentiation function.}
    \label{fig:expon}
\end{figure}

\begin{figure}[tbp]
    \centering
    \includegraphics[width=0.4\textwidth, trim=10cm 2.5cm 7cm 4cm, clip]{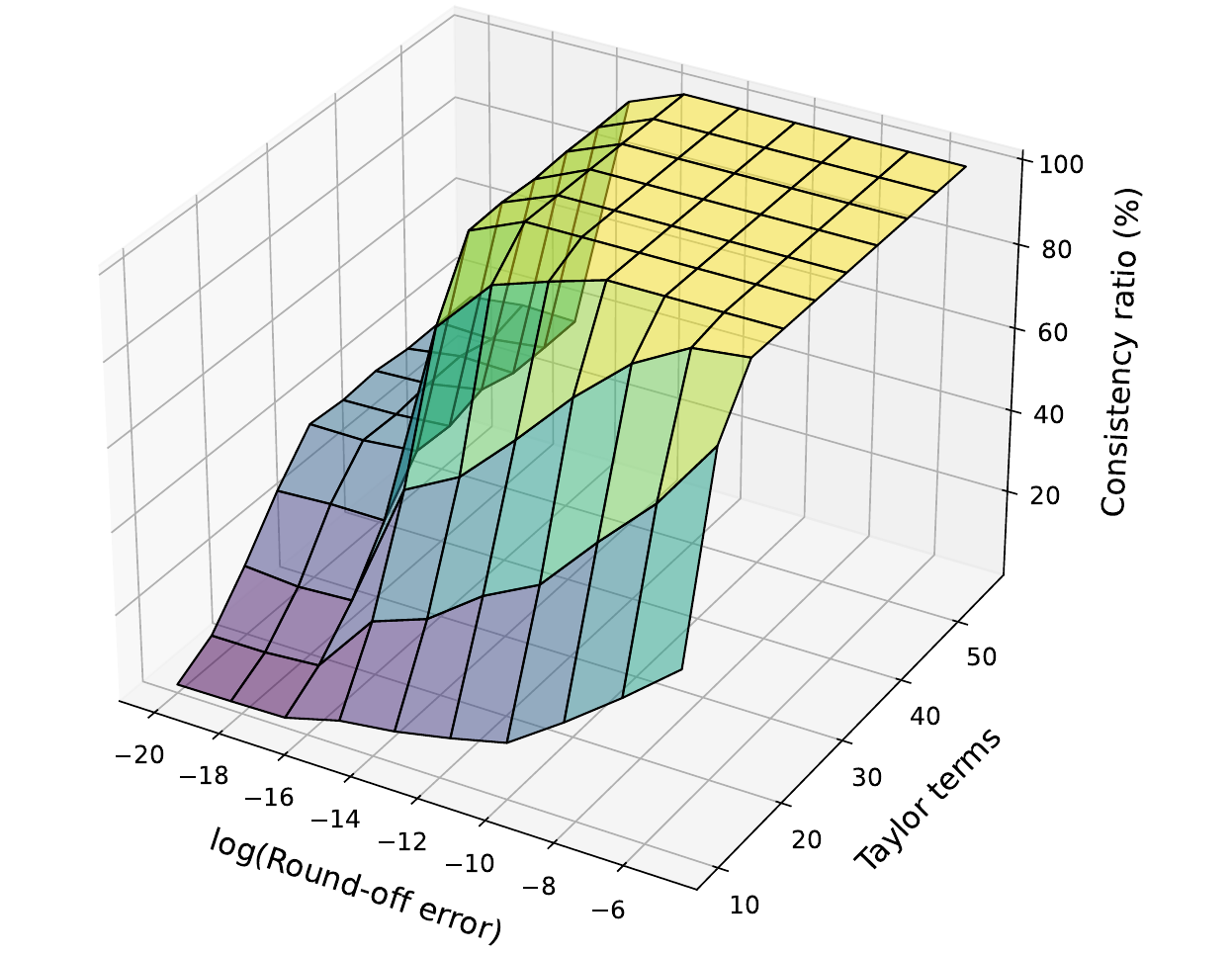}
    \caption{Effect of round-off error and number of Taylor terms to the accuracy of the proposed architecture for sigmoid function.}
    \label{fig:sigmoid}
\end{figure}

\begin{figure}[tbp]
    \centering
    \includegraphics[width=0.4\textwidth, trim=10cm 2.5cm 7cm 4cm, clip]{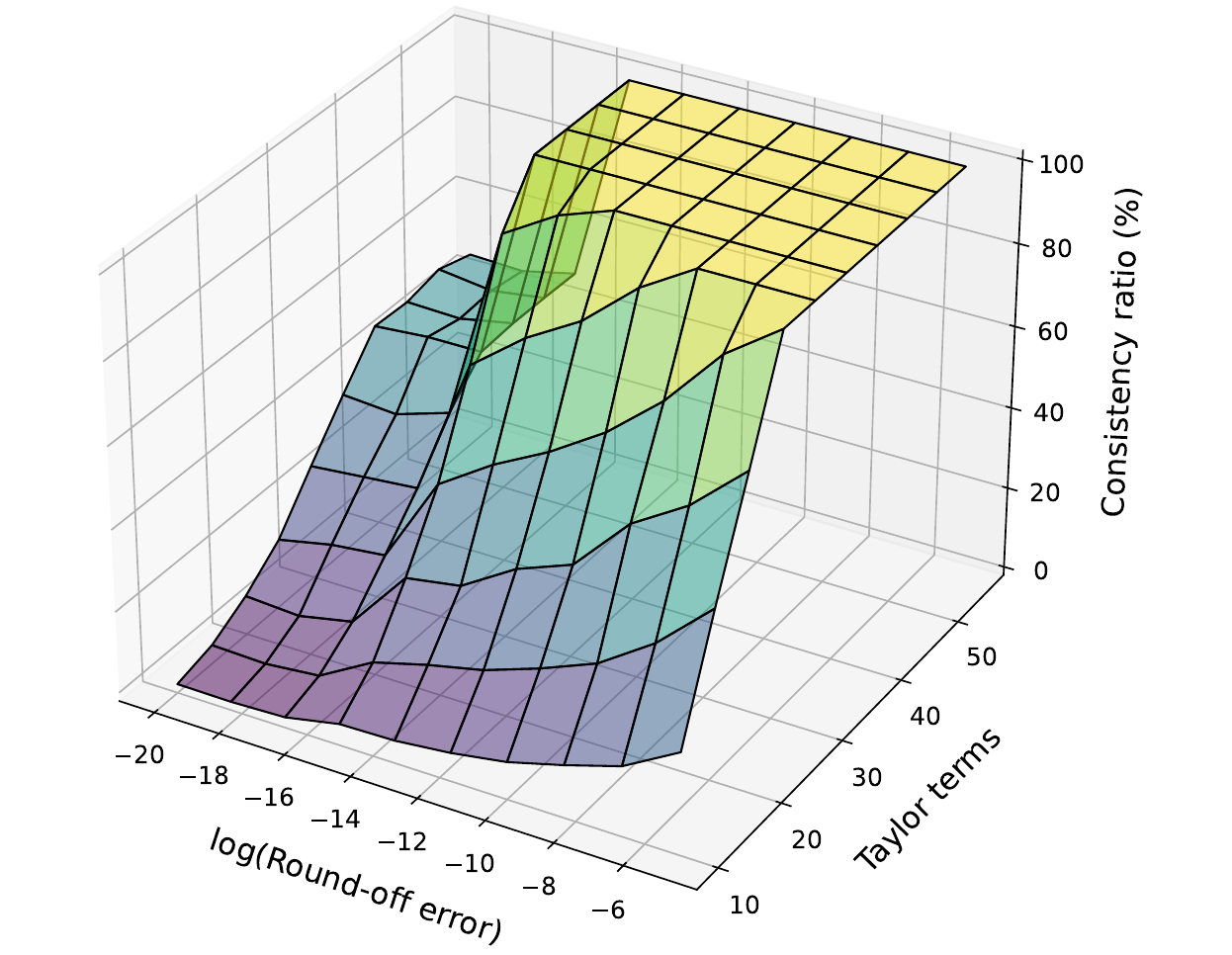}
    \caption{Effect of round-off error and number of Taylor terms to the accuracy of the proposed architecture for tanh function.}
    \label{fig:tanh}
\end{figure}

We implement the exponentiation function using a Taylor expansion and use it to compute the sigmoid and tanh functions. Moreover, when dealing with floating-point arithmetic, round-off errors must be taken into account. This means that even if no faults are injected, the equation checks may not match, causing the design to incorrectly declare a fault due to round-off errors. Consequently, a trade-off between accuracy and overhead must be considered. Figs. \ref{fig:expon}, \ref{fig:sigmoid}, and \ref{fig:tanh} illustrate these trade-offs for the exponentiation, sigmoid, and tanh functions, respectively, for the input range of $[-3, 3]$. Specifically, consistency ratio in these figures show how the number of Taylor terms and the round-off error affect whether the fault-detection architecture incorrectly interprets round-off errors as injected faults when no actual faults are present. 
The combination of round-off error and number of terms that achieves a 100\% consistency ratio is chosen according to the computational capabilities of the target platform. Fewer terms are preferable for resource-constrained devices, while applications requiring higher accuracy demand finer round-off error precision.
Based on these figures, we have selected different parameter settings and round-off error thresholds for each function. Moreover, we observed that many of the injected faults had little to no effect on the output, particularly those affecting the lower bits of the significant in the IEEE format. The selected parameter settings are summarized in Table~\ref{table:settings}.

\begin{table}[t]
\label{sim-settings}
\begin{centering}
\caption{Simulation settings}
\label{table:settings}
\par\end{centering}
\centering{}\centering%
\begin{tabular}{cccc}
\toprule 
\multirow{2}{*}{Parameter} & \multicolumn{3}{c}{Simulation Settings}\tabularnewline
\cmidrule{2-4}
 & expo & sigmoid & tanh\tabularnewline
\midrule
\midrule 
Input signal & \multicolumn{3}{c}{Range of $[-3,3]$}\tabularnewline
Iterations for each simulation & \multicolumn{3}{c}{1000}\tabularnewline
Taylor terms number & 30 & 30 & 40\tabularnewline

Round-off error & $10^{-14}$ & $10^{-14}$ & $10^{-15}$\tabularnewline
\bottomrule
\end{tabular}
\end{table}

\begin{table}[t]
\caption{Our comprehensive fault detection simulation results For ``Random'' injection type}
\label{error_big1}
\centering{}\centering%
\begin{tabular}{cccccc}
\toprule 
\multirow{2}{*}{Model} & \multicolumn{2}{c}{\#} & \multicolumn{3}{c}{Fault Detection Ratio (\%)}\tabularnewline
\cmidrule{2-6}
 & $n$ & $m$ & Expo & Sigmoid & tanh\tabularnewline
\midrule
\midrule 
 & 1 & 1 & 96.1 & 99.1 & 96.5\tabularnewline
\multicolumn{1}{c}{} & 1 & 5 & 96.6 & 99.4 & 97.7\tabularnewline
\multirow{2}{*}{Bit flipping} & 3 & 1 & 95.6 & 98.2 & 95.3\tabularnewline
 & 3 & 5 & 98.3 & 99.5 & 97.3\tabularnewline
 & 6 & 1 & 97.5 & 98.7 & 97.7\tabularnewline
 & 6 & 5 & 99.5 & 99.9 & 99.2\tabularnewline
\midrule
 & 1 & 1 & 98.2 & 99.2 & 96.6\tabularnewline
 & 1 & 5 & 98.2 & 98.9 & 97.8\tabularnewline
\multirow{2}{*}{Stuck-at-1} & 3 & 1 & 95.9 & 98.8 & 95.1\tabularnewline
 & 3 & 5 & 97 & 98.3 & 96.9\tabularnewline
 & 6 & 1 & 95.9 & 98.7 & 94.8\tabularnewline
 & 6 & 5 & 98.6 & 99.6 & 99\tabularnewline
\midrule
 & 1 & 1 & 98.1 & 99.1 & 97.1\tabularnewline
 & 1 & 5 & 97.4 & 98.8 & 95.7\tabularnewline
\multirow{2}{*}{Stuck-at-0} & 3 & 1 & 96.8 & 98.5 & 94\tabularnewline
 & 3 & 5 & 97.7 & 98.7 & 96.9\tabularnewline
 & 6 & 1 & 96.2 & 98 & 94.5\tabularnewline
 & 6 & 5 & 98.8 & 99.8 & 98.5\tabularnewline
\bottomrule
\end{tabular}
\end{table}

\begin{table}[t]
\caption{Our comprehensive fault detection simulation results For ``Burst'' injection type}
\label{error_big2}
\centering
\begin{tabular}{cccccc}
\toprule 
\multirow{2}{*}{Model} & \multicolumn{2}{c}{\#} & \multicolumn{3}{c}{Fault Detection Ratio (\%)}\tabularnewline
\cmidrule{2-6}
 & $n$ & $m$ & Expo & Sigmoid & tanh\tabularnewline
\midrule
\midrule 
 & 1 & 2 & 95.2 & 97.9 & 96\tabularnewline
 & 1 & 5 & 95.9 & 98 & 95.8\tabularnewline
\multirow{2}{*}{Bit flipping} & 3 & 2 & 94.2 & 98 & 96.1\tabularnewline
 & 3 & 5 & 95 & 98.2 & 95.6\tabularnewline
 & 6 & 2 & 97.1 & 98.3 & 97.3\tabularnewline
 & 6 & 5 & 97.8 & 98.1 & 98.3\tabularnewline
\midrule
 & 1 & 2 & 97.8 & 98.4 & 95.9\tabularnewline
 & 1 & 5 & 95.3 & 98.3 & 97.3\tabularnewline
\multirow{2}{*}{Stuck-at-1} & 3 & 2 & 94.4 & 98.3 & 95.1\tabularnewline
 & 3 & 5 & 94.9 & 98.4 & 95.5\tabularnewline
 & 6 & 2 & 96.1 & 98.2 & 96\tabularnewline
 & 6 & 5 & 96.3 & 98.3 & 96.8\tabularnewline
\midrule
 & 1 & 2 & 98.5 & 98.4 & 96.8\tabularnewline
 & 1 & 5 & 95 & 97.9 & 95.2\tabularnewline
\multirow{2}{*}{Stuck-at-0} & 3 & 2 & 95.3 & 98.2 & 94.7\tabularnewline
 & 3 & 5 & 94.9 & 97 & 95.4\tabularnewline
 & 6 & 2 & 96.6 & 98.3 & 94.3\tabularnewline
 & 6 & 5 & 96 & 98.8 & 98.1\tabularnewline
\bottomrule
\end{tabular}
\end{table}

\begin{table}[t]
\caption{Our comprehensive fault detection simulation results for ``Skipping'' and ``Total Random'' models}
\label{error_small}
\centering{}\centering%
\begin{tabular}{ccccc}
\toprule 
\multirow{2}{*}{Model} & \multicolumn{1}{c}{\#} & \multicolumn{3}{c}{Fault Detection Ratio (\%)}\tabularnewline
\cmidrule{2-5}
 & $n$ & Expo & sigmoid & tanh\tabularnewline
\midrule
\midrule 
 & 1 & 90.8 & 93 & 92.9\tabularnewline
\multicolumn{1}{c}{} & 2 & 95.5 & 95.4 & 95.9\tabularnewline
\multirow{2}{*}{Skipping} & 3 & 96.8 & 97.1 & 97.5\tabularnewline
 & 4 & 98.1 & 98.7 & 98.4\tabularnewline
 & 5 & 98.5 & 99.1 & 98.4\tabularnewline
 & 6 & 99.5 & 99.7 & 99.4\tabularnewline
\midrule
 & 1 & 89.1 & 92 & 94.6\tabularnewline
 & 2 & 94.8 & 94.1 & 96.2\tabularnewline
\multirow{2}{*}{Total Random} & 3 & 96.3 & 97.1 & 98\tabularnewline
 & 4 & 98.6 & 97.7 & 98\tabularnewline
 & 5 & 99.2 & 98.6 & 99.2\tabularnewline
 & 6 & 99.7 & 99.5 & 99.2\tabularnewline
\bottomrule
\end{tabular}
\end{table}

\subsubsection{Number of Fault Injections}
For the number of injections, we consider two parameters, $n$ and $m$. Here, $n$ specifies the number of faulty terms, and $m$ indicates how many bits of each selected faulty term are going to be corrupted.

\subsubsection{Injection Types}
The injection types illustrate how the terms and bits to be corrupted are selected. We consider two different types of fault injection: Random and Burst. In the Random type, both the faulty terms and the faulty bits of each term are selected randomly. In the Burst type, after selecting the faulty terms randomly, only the first faulty bit is chosen randomly, and the following $m$ consecutive bits are selected to be corrupted.

\subsubsection{Injection Models}
After selecting the faulty terms and bits, the injection model specifies how these bits will be effected. We consider five different models: Bit Flipping, Stuck-at-0, Stuck-at-1, Instruction Skipping, and Total Random. In the Bit Flipping model, the selected bits are inverted (XOR with 1). In the Stuck-at-1 and Stuck-at-0 models, the bits are set to 1 (OR with 1) or 0 (AND with 0), respectively. Instruction Skipping means that the computation of an entire term is skipped, and Total Random means the faulty term is replaced with a completely new random number.

\subsection{Error Coverage Ratio}
Tables \ref{error_big1}, \ref{error_big2} and \ref{error_small} illustrate the detection ratio of the proposed method under the disussed fault-injection scenarios and varying numbers of injections. These simulations are performed based on the settings summarized in Table~\ref{table:settings}.

\section{Hardware and Software Implementation}

\begin{table}[t]
\centering
\caption{Comparative analysis of inference latency and memory footprint for baseline and protected activation functions in TensorFlow}\vspace{1mm}
\label{tab:inference_tensor}
\resizebox{\columnwidth}{!}{%
\begin{tabular}{ccc}
\toprule
Activation Function & \begin{tabular}[c]{@{}c@{}}Average Inference\\ Time (ms)\end{tabular} & Model Size (MB) \\ \hline
tanh\textsuperscript{1}          & 0.284 & 0.37 \\ \hline
error\_tanh\textsuperscript{2}    & 0.432 & 0.37 \\ \hline
sigmoid\textsuperscript{1}       & 0.294 & 0.37 \\ \hline
error\_sigmoid\textsuperscript{2} & 0.458 & 0.37 \\ \hline
ReLu\textsuperscript{1}          & 0.283 & 0.37 \\ \hline
error\_ReLu\textsuperscript{2}    & 0.313 & 0.37 \\ \hline
\multicolumn{3}{l}{\textsuperscript{1} Baseline, \textsuperscript{2} Protected}
\end{tabular}%
}
\end{table}
As proposed, the error detection schemes target the activation function. 
We evaluate the overhead of these schemes in two settings: integrated within a full model and as standalone implementations. 
In the software domain, the proposed error detection mechanism is integrated into TensorFlow 2.x, leveraging static graph compilation. In contrast, for microcontroller and FPGA platforms, we implement the baseline activation function and its error-detecting counterpart as standalone modules, without embedding them into a complete model, and measure the associated overhead. 

\subsection{Software Framework Integration }
We validated the integration of the proposed error detection mechanism using a DNN trained on the NASA Turbofan Jet Engine dataset~\cite{NasaDataset}. Our presented error-detection-enabled DNN model was trained and evaluated on an NVIDIA GeForce RTX 3070 Ti Laptop GPU, operating at a 210 MHz clock speed and equipped with 8 GB of dedicated memory.
This software-based implementation serves as a proof-of-concept for the scheme's applicability in real-world models. Given that the performance impact in the software domain is minimal, the subsequent evaluation of overhead is restricted to the standalone hardware modules implemented on FPGA and microcontroller platforms. 
This dataset and model were chosen because prognostics and health management play a critical role in industry, helping predict the condition of assets to minimize downtime and prevent failures.
The occurrence of either fault attacks or natural faults during the inference phase of a DNN model can result in catastrophic consequences for this industry.
In this dataset, the primary objective is to estimate the remaining useful life (RUL) of each engine in the test set, where RUL represents the number of operational cycles (or flights) an engine has left after the final recorded data point.\\
We designed a DNN to classify the degradation levels of the monitored equipment into six categories corresponding to different ranges of RUL.
In our terminology, this model is referred to as the baseline model, as it does not incorporate any fault detection scheme.
The input to the model consists of multivariate time-series sensor readings collected from the equipment.
The model architecture starts with a 1D Convolutional Neural Network followed by a max-pooling operation to extract local temporal patterns and reduce noise. 
The convolution layer enables the network to capture short-term dependencies and localized trends in the sensor data, which are often indicative of early stage degradation.
The extracted features are then passed through several fully connected layers to perform high-level feature abstraction and nonlinear combination of the learned representations. 
The final dense layer with six neurons and a softmax activation outputs the probability distribution over six RUL classes.
Although more complex deep learning architectures such as Long Short-Term Memory (LSTM) networks could potentially achieve higher prediction accuracy by capturing long-term temporal dependencies in sensor data, they typically require significantly more computational and memory resources.
In contrast, our goal was to design a lightweight model that can be efficiently deployed on resource-constrained embedded systems, where computational power and memory availability are limited.
Therefore, we adopted a 1D Convolutional Neural Network combined with fully connected layers, which provides a good balance between performance and efficiency.
This design choice allows us to evaluate the overhead introduced by our proposed fault detection scheme on embedded hardware without the burden of high model complexity associated with recurrent architectures like LSTMs.
\\
We integrated the proposed error detection scheme directly into TensorFlow~2.x by embedding it within a high-fidelity numerical approximation of the activation functions. The error detection logic was incorporated into the TensorFlow computation graph to ensure tight coupling with model execution. To enable efficient and deterministic inference, the evaluation routine was annotated with the \texttt{@tf.function} decorator, thereby triggering static graph compilation and allowing the error detection operations to be optimized jointly with the baseline activation function.

The Taylor series approximation was restructured to support the proposed error detection mechanism by precomputing inverse factorial coefficients and storing them as constant tensors. This approach eliminates redundant arithmetic operations during execution and enables TensorFlow to exploit constant folding and static loop unrolling during the graph tracing phase. To preserve numerical stability and prevent spurious error detections caused by divergence, an input clipping layer was integrated into the computational graph, constraining inputs to the convergence interval of the Taylor series.
Table~\ref{tab:inference_tensor} summarizes the resulting inference latency over 1,000 samples and model size for activation functions with and without the proposed error detection scheme.

The purpose of developing multiple models employing different activation functions, both with and without integrated error detection, was not to optimize predictive accuracy.
Rather, the primary objective was to evaluate the overall overhead introduced at the model level when error detection mechanisms are enabled.
To provide context for the model performance, the evaluation on the test dataset showed that the best-performing configuration achieved an accuracy and recall (sensitivity) of 0.8246 with an F1-score of 0.8341, while the lowest-performing configuration resulted in an accuracy and recall of 0.7143 and an F1-score of 0.7401.

\subsection{Microcontroller Implementation}

To evaluate the proposed error detection scheme in terms of clock-cycle performance on microcontrollers, we conducted experiments on an Arduino Uno equipped with an ATmega328P microcontroller.
Table~\ref{microcontroller} presents the clock cycles of the baseline activation functions and their protected counterparts. It should be noted that, since the ReLu activation employs a recomputation-based approach, it incurs a 100\% overhead on microcontroller based architectures.

The \texttt{sigmoid\_error\_detection} function serves as the primary controller for the fault-tolerant sigmoid calculation, integrating input normalization, series expansion, and a mathematical consistency check. After clipping the input to a stable range, it generates Maclaurin series terms to compute the sigmoid value $y$. The function's critical feature is its error-detection logic, which validates the result by checking if the identity $\frac{y}{1-y} = e^x$ holds true within a specified precision threshold ($\epsilon = 10^{-14}$). If the calculated ratio deviates from the independently computed exponential value, the function flags a hardware or computational error.
The \texttt{Maclaurin\_terms} function implements an efficient iterative algorithm to compute the components of the power series for $e^x$. To minimize computational overhead on resource-constrained hardware, it utilizes the recurrence relation $T_i = T_{i-1} \cdot \frac{x}{i}$, where $T_i$ represents the $i$-th term of the series. This approach reduces the complexity from $O(n^2)$ to $O(n)$, making high-order approximations feasible for 8-bit microcontrollers.
The \texttt{clip\_by\_value} function performs input saturation to ensure numerical stability during series expansion. By constraining the input variable $x$ to the interval $[-3, 3]$, it prevents the exponential terms from exceeding the dynamic range of the double data type. This bounding is critical for maintaining the convergence properties of the Maclaurin series and avoiding floating-point overflow during the subsequent error-detection phase.
\begin{table}[t]
\caption{Evaluation of clock cycles for baseline and protected activation functions on ATmega328P microcontroller (16 MHz)}\vspace{0.5mm}
\label{microcontroller}
\resizebox{\columnwidth}{!}{%
\begin{tabular}{cccc}
\toprule
Activation Function & Clock Cycles & Time ($\mu$s) & Overhead (\%) \\ \hline
tanh\textsuperscript{1}          & 9,933,888 & 620,868 & -       \\ \hline
error\_tanh\textsuperscript{2}    & 9,934,400 & 620,900 & 0.0052\% \\ \hline
sigmoid\textsuperscript{1}       & 9,927,808 & 620,488 & -       \\ \hline
error\_sigmoid\textsuperscript{2} & 9,940,928 & 621,308 & 0.132\%  \\ \hline
\multicolumn{4}{l}{\textsuperscript{1} Baseline, \textsuperscript{2} Protected}
\end{tabular}%
}
\end{table}

\subsection{FPGA Implementation}

\begin{table}[t]
\centering
\caption{AMD/Xilinx Artix-7, xc7a200tfbg676-2 hardware implementation results for sigmoid function}
\label{fpga-fig}
\resizebox{\columnwidth}{!}{%
\begin{tabular}{cccc}
\toprule
Scheme                                  &     & \begin{tabular}[c]{@{}c@{}}Protected\end{tabular} & \begin{tabular}[c]{@{}c@{}}Baseline\end{tabular} \\ \hline
\multirow{2}{*}{Area}                   & LUTs      & 2,934      & 2,896      \\ \cline{2-4} 
                                        & DSPs      & 18         & 18         \\ \hline
\multirow{2}{*}{Timing}                 & \begin{tabular}[c]{@{}c@{}}Latency (CCs)\end{tabular}   & 9          &  7         \\ \cline{2-4} 
                                        & \begin{tabular}[c]{@{}c@{}}Total time (ns)\end{tabular} & 64.26      & 49.98      \\ \hline
\begin{tabular}[c]{@{}c@{}}Power (W)\\ @ 140 MHz\end{tabular} &     & 0.214      & 0.212      \\ \hline
Energy (nJ)                             &     & 13.75      & 10.59      \\ \hline
\end{tabular}%
}
\end{table}

To evaluate the overhead, we focus on the sigmoid function in our FPGA implementation, since tanh exhibits similar behavior and can be realized using largely the same submodules.
The high computational cost of the sigmoid function poses significant challenges for real-time applications, especially on resource-constrained hardware or in large-scale networks.
To mitigate these issues, several approximation techniques, such as Taylor series expansion and piecewise linear approximation, are commonly used in the digital implementation of nonlinear activation functions~\cite{approx_1,approx_2}.
The selection of an appropriate approximation method depends on factors including required accuracy, computational constraints, available hardware resources, and application-specific requirements. 
Each approach entails trade-offs among accuracy, execution speed, and hardware complexity, making careful evaluation essential for informed design choices.
Our primary objective in the FPGA implementation is to demonstrate the overhead introduced by the proposed scheme relative to the baseline implementation, rather than to develop an optimized FPGA design.

Figure~4 illustrates the hardware architecture of the proposed design.
The architecture comprises three main modules: floating-point addition/subtraction, 
floating-point multiplication, and floating-point division. 
Each term of the Taylor series expansion is computed in a single clock cycle 
within the term calculation sub-block shown in Figure~4. 
A total of 5 terms are used in the approximation, with each term computed in one clock cycle, resulting in 5 clock cycles for term evaluation. The accumulation of these terms is performed in a single clock cycle using the floating-point addition module. The final sigmoid output, computed as \( \frac{1}{1 + e^{-x}} \), is obtained with 1 additional clock cycle using the floating-point division module. Consequently, the sigmoid function requires a total of 7 clock cycles.
For error detection, as terms are pre-calculated, the computation of \( e^{x} \) requires one additional floating-point addition, followed by a comparison between the sigmoid output \( y \) and the computed value of \( e^{x} \), which together incur 2 extra clock cycles. Therefore, the complete system, including error detection, operates in a total of 9 clock cycles. The comparison is performed using an appropriate round-off error corresponding to the number of terms, as determined by the simulation results described in Section 4.
Table~\ref{fpga-fig} presents the hardware implementation results for the proposed error detection scheme alongside the baseline implementation without error detection.
The design prioritizes increased latency rather than additional hardware resources; consequently, the proposed scheme incurs nearly zero area overhead while approximately doubling the execution latency.

\begin{figure*}[!t]
    \centering
    \includegraphics[width=\textwidth]{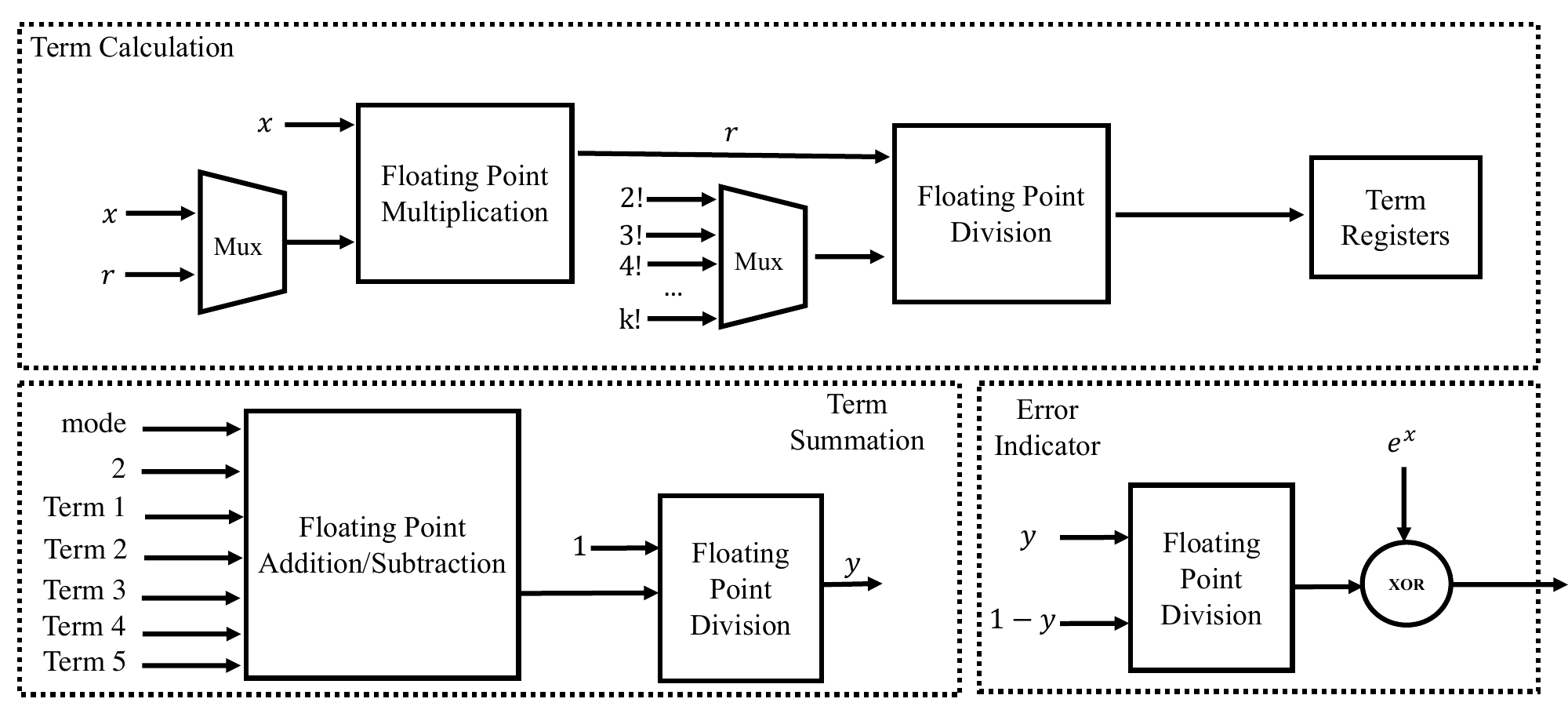}
    \caption{Hardware Architecture of Sigmoid Function with Error Detection.}
    \label{fig:exponentiation}
\end{figure*}

\subsubsection{Floating-Point Multiplication}
The floating-point multiplier realizes IEEE 754 single-precision multiplication using fully combinational logic, enabling completion within a single clock cycle.
The sign of the result is determined via an XOR operation on the operand sign bits, while the exponent is computed by summing the input exponents followed by bias adjustment.
The mantissas, treated as 24-bit significands that include the implicit leading bit, are multiplied to generate a 48-bit intermediate result.
This product is then normalized according to the presence of a carry-out.
Rounding is performed by evaluating the lower 23 bits, after which truncation yields the final mantissa.
The output is assembled in IEEE 754 format by concatenating the calculated sign bit, the 8-bit exponent, and the 23-bit rounded mantissa, with special cases handled through appropriate exception encoding.
\subsubsection{Floating-Point Division}
The divider employs the Newton-Raphson method to compute floating-point division through iterative reciprocal approximation. 
Rather than performing long division, the design normalizes the divisor to a standard range by exponent adjustment, then generates an initial reciprocal estimate using a precomputed constant multiplied by the normalized divisor.
This approximation is progressively refined through successive iterations using the Newton-Raphson update formula, with each iteration leveraging the multiplier and adder modules to converge toward the true reciprocal.
\subsubsection{Floating-Point Addition/Subtraction}
Our design includes two types of floating-point addition/subtraction units.
The first unit performs addition or subtraction on a pair of inputs and is utilized within the division module.
The second unit is responsible for summing five terms together with a constant integer (either 1 or 2, resulting in a total of 6 operands), depending on the selected mode. 
This unit operates in two modes: when the mode is set to 0, it sums all terms to compute $e^{x}$; when the mode is set to 1, it negates one of the intermediate inputs to compute $e^{-x} + 1$. 
All operations are completed within a single clock cycle.

\section{Discussion}
\subsection{Comparision With Other Countermeasures}
Control-Flow Integrity (CFI)~\cite{cfi} is a commonly used software countermeasure designed to prevent control-flow hijacking by enforcing adherence to a statically derived Control-Flow Graph (CFG).
The CFG captures all valid execution paths of a program, modeling basic blocks as nodes and legitimate control transfers, such as branches, function calls, and returns, as edges. To enforce this model, CFI introduces runtime checks that ensure each control transfer targets only the allowed successors specified by the CFG.
When applied to microcontrollers, CFI can detect fault attacks, including voltage or clock glitching, that result in instruction skipping or unintended jumps to addresses outside the valid control-flow structure.
Nevertheless, faults that redirect execution to alternative but still permissible nodes within the CFG may evade detection, particularly in coarse-grained CFI schemes favored in resource-constrained embedded environments.
As a result, although CFI effectively mitigates illegal control-flow deviations, it does not fully protect against data corruption or fault-induced deviations that remain within the CFG, underscoring the necessity of integrating complementary algorithm-level or data-flow protection techniques for robust fault resilience.
Furthermore, CFI is applicable only to instruction-based processors; it cannot be directly employed in hardware implementations or FPGA platforms. In such cases, algorithm-level error detection or redundancy-based techniques must be adopted instead.

Existing fault detection mechanisms for DNNs adopt fundamentally different strategies to address hardware-induced faults and adversarial fault injection attacks.
ThresholdChecking~\cite{rangedefence} relies on a lightweight anomaly detection approach that monitors activation values against pre-characterized bounds. While this method introduces negligible computational overhead, its detection capability is limited, typically identifying less than 10\% of faults under modern, coordinated attack scenarios.
DeepDyve~\cite{deepdyve}, in contrast, performs dynamic verification by deploying an auxiliary checker neural network that approximates the primary model and flags discrepancies through end-to-end output comparison. Although this approach achieves high detection rates (75-99\%), it incurs significant runtime overhead (6-37\%) and requires maintaining two separate models, making it less suitable for resource-constrained platforms.

\textit{MAED} adopts a complementary design philosophy tailored specifically for embedded and edge AI systems that are vulnerable to physical fault injection attacks. Instead of monitoring output anomalies or relying on model redundancy, \textit{MAED} exploits the intrinsic mathematical identities of activation functions to perform algorithm-level fault detection during inference. This enables near-complete error coverage while introducing minimal overhead on constrained hardware, including less than 1\% clock-cycle overhead on microcontrollers and negligible area overhead on FPGAs. Unlike ThresholdChecking, which is largely ineffective for quantized DNNs and sophisticated attacks such as BitFlip, and DeepDyve, whose latency is prohibitive for real-time embedded deployment, \textit{MAED} is uniquely effective against practical physical attacks (e.g., DeepLaser) that manipulate control flow and arithmetic operations within activation functions. 
These complementary characteristics suggest that robust protection for safety-critical DNN systems can be achieved by integrating \textit{MAED}'s lightweight mitigation against physical fault attacks with higher-overhead defense mechanisms, such as DeepDyve, in environments with fewer resource constraints, including cloud and edge platforms. This combined approach enables a balanced trade-off between security strength and implementation cost across diverse deployment scenarios.

\subsection{Limitation of Proposed Schemes}
The \textit{MAED} defense framework is purposefully designed to detect and mitigate fault injection attacks specifically localized within activation function layers. Although its primary scope does not encompass the direct protection of DNN weights, \textit{MAED} demonstrates significant robustness against faults that manifest as memory level bit flips within the inputs to activation functions. This specialized detection capability ensures its practical utility across diverse deployment scenarios, ranging from high performance cloud infrastructures to embedded environments. A core principle of the \textit{MAED} architecture is its design optimization for resource-constrained embedded systems, where power and computational overhead are critical design factors. Furthermore, when deployed in conjunction with complementary weight protection protocols, \textit{MAED} facilitates a multi layered security architecture. This collaborative approach provides a comprehensive defense strategy, significantly hardening embedded DNNs against a broad spectrum of fault injection attacks.
\section{Conclusions}
We present \textit{MAED}, which, to the best of our knowledge, represents the first comprehensive resilience study that explicitly focuses on the algorithm-level activation functions utilized in DNNs during the inference phase. Prior works have largely examined fault attacks at the hardware or system level, but there has been no systematic investigation into how faults affect the computations within activation functions themselves and how such faults can propagate through the network. Our study aims to fill this gap by proposing methods to detect and mitigate fault attacks specifically targeting these crucial nonlinear components in DNNs, with a focus on embedded and resource constrained systems.  This work provides fundamental insights into the role of activation functions in DNN resilience and demonstrates how incorporating lightweight security checks directly within these functions can significantly reduce the impact of fault injections during inference. By addressing errors at the algorithmic level, \textit{MAED} offers a novel perspective on safeguarding DNN computations beyond conventional hardware- or software-centric fault detection mechanisms. Furthermore, we propose efficient and lightweight mitigation techniques that are suitable for both hardware and software implementations, and we systematically evaluate the overhead introduced by integrating our scheme with a baseline system (i.e., without error detection). This evaluation is performed on an AMD/Xilinx Artix-7 FPGA, an ATmega328P microcontroller, and through integration with TensorFlow. Our results indicate that on the microcontroller, the proposed error detection introduces less than 1\% additional clock cycle overhead, while on the FPGA it incurs almost no extra area, with a trade-off of approximately 20\% increased latency for sigmoid and tanh activation functions.  

% As future work, we plan to extend our study to encompass a wider range of fault injection attacks on DNNs and investigate additional network components beyond activation functions, further broadening the applicability of algorithm-level fault detection in secure and robust deep learning systems.

% \input{sections/openscience}
% \input{sections/ethics}

% optional clearing of the page
{ 
\bibliographystyle{IEEEtran}
\bibliography{references}}
\begin{IEEEbiography}[{{\includegraphics[scale=0.07]{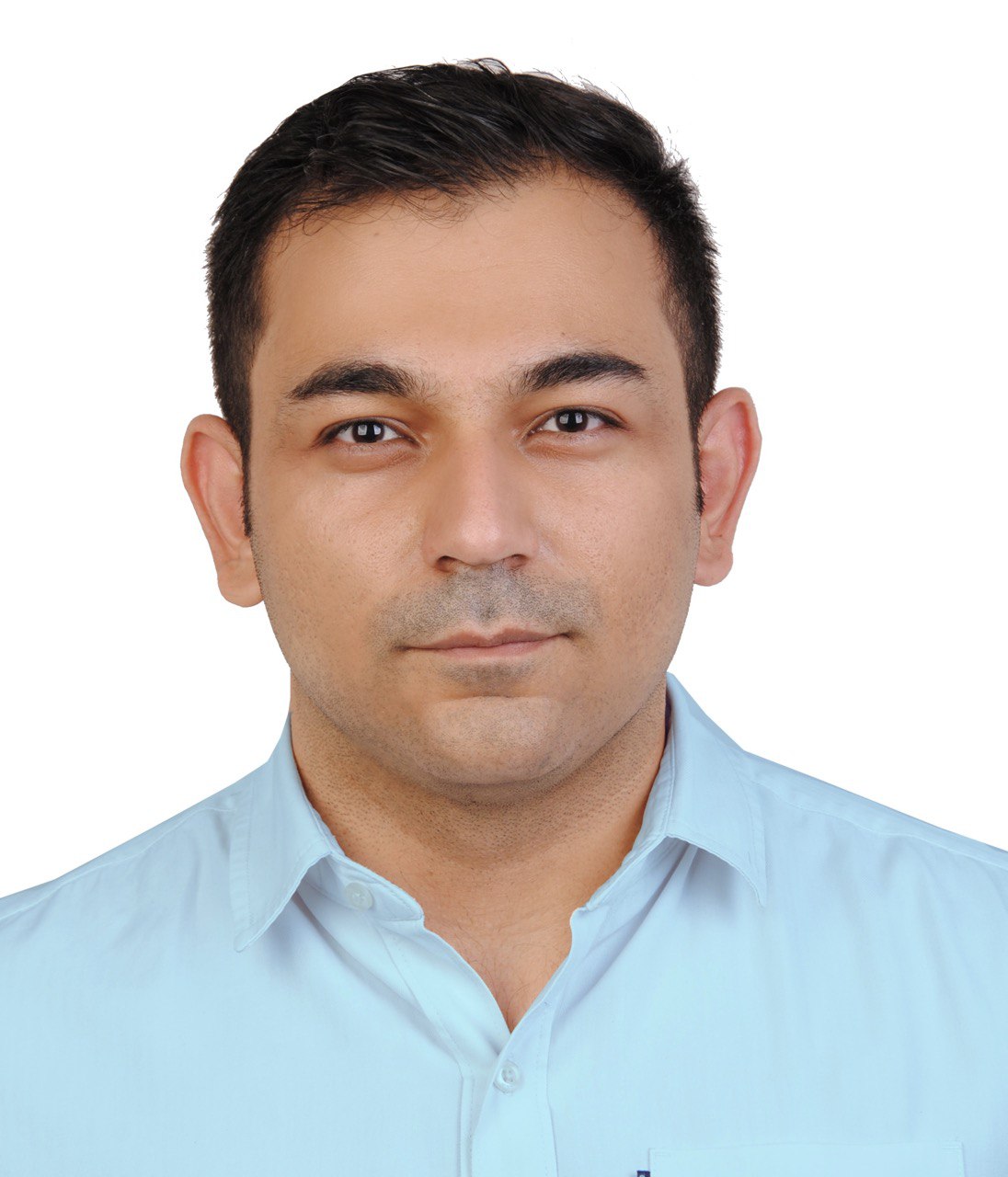}}}]{}
Kasra Ahmadi received his B.Sc. degree in Computer Engineering from
Isfahan University of Technology, Isfahan, Iran and his M.Sc. degree
in Information Technology from AmirKabir University of Technology,
Tehran, Iran. He is currently a Ph.D. candidate within the Bellini College of Artificial Intelligence, Cybersecurity, and Computing at the University of South Florida.
His current research interests include machine learning security, differential-privacy, algorithm-level error detection for cryptographic embedded systems, post-quantum cryptography and side-channel
attacks.
\end{IEEEbiography}
\begin{IEEEbiography}[{{\includegraphics[scale=0.43]{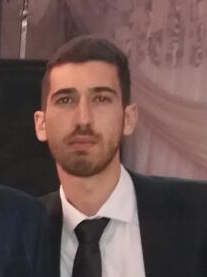}}}]{}
Saeed Aghapour received his B.Sc. degree in Electrical Engineering,
from Babol Noshirvani University of Technology, Babol, Iran in 2014
and his M.Sc. in Communication Cryptology from the Electrical Engineering
department of Sharif University of technology, Tehran, Iran in 2016.
He is currently
Ph.D. candidate at the Bellini College of Artificial Intelligence, Cybersecurity and Computing
at the University of South Florida.
His current research interests include applied cryptography, post-quantum
cryptography and their secure implementations, hardware security,
fault injection analysis, and security for IoT, smart grid and embedded
systems.
\end{IEEEbiography}
\begin{IEEEbiography}[{{\includegraphics[scale=0.09]{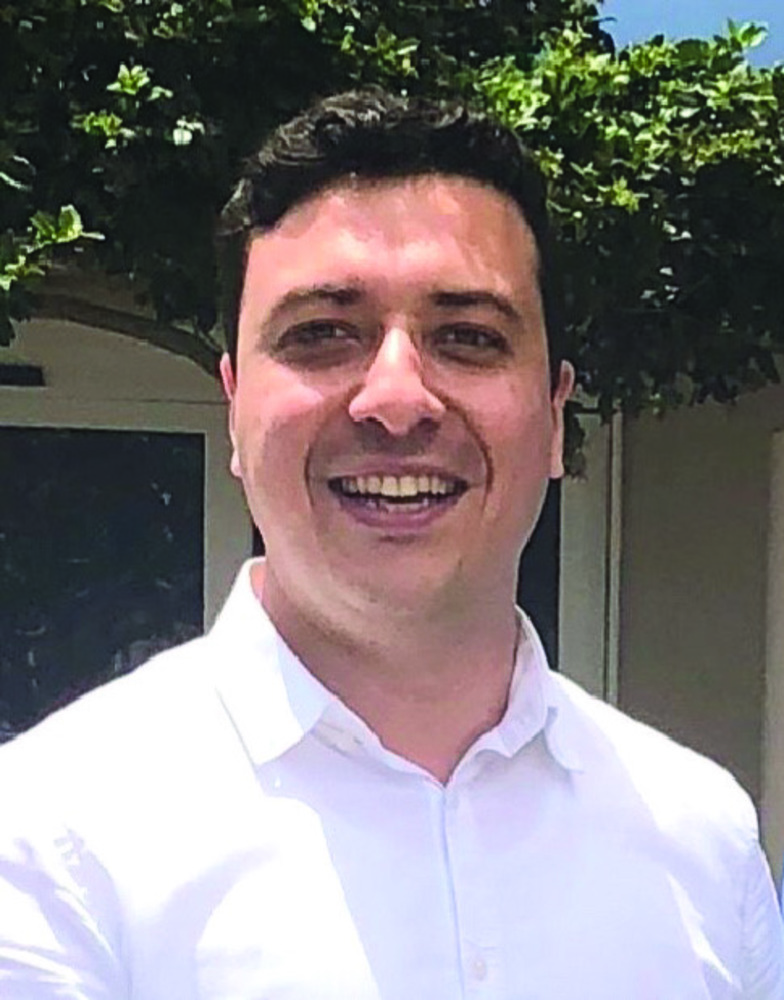}}}]{}
Mehran Mozaffari Kermani (S'00-M'11-SM'16) received the B.Sc. degree
from the University of Tehran, Tehran, Iran, in 2005, and the M.E.Sc.
and Ph.D. degrees from University of Western Ontario, London, Canada,
in 2007 and 2011, respectively. In 2012, he joined the Electrical
Engineering Department, Princeton University, New Jersey, as an NSERC
post-doctoral research fellow. From 2013-2017 he was a faculty with
Rochester Institute of Technology and starting 2017, he is an Associate
Professor with the Bellini College of
Artificial Intelligence, Cybersecurity, and Computing at the University of South Florida. He has served as an Associate
Editor for the\emph{ IEEE Transactions on VLSI Systems}, the \emph{ACM
Transactions on Embedded Computing Systems}, and the \emph{IEEE Transactions
on Circuits and Systems. }He has been the TPC member for HOST (Publications
Chair), CCS (Publications Chair), DAC, DATE, RFIDSec, LightSec, WAIFI,
FDTC, and DFT. He was a recipient of the prestigious Natural Sciences
and Engineering Research Council of Canada Post-Doctoral Research
Fellowship in 2011 and the Texas Instruments Faculty Award (Douglas
Harvey) in 2014. He is also the awardee for USF 2021 Faculty Outstanding
Research Achievement Award, and USF College of Engineering's 2018
Outstanding Junior Research Achievement Award. He is a Senior Member
of the IEEE.
\end{IEEEbiography}
\begin{IEEEbiography}[{{\includegraphics[scale=0.09]{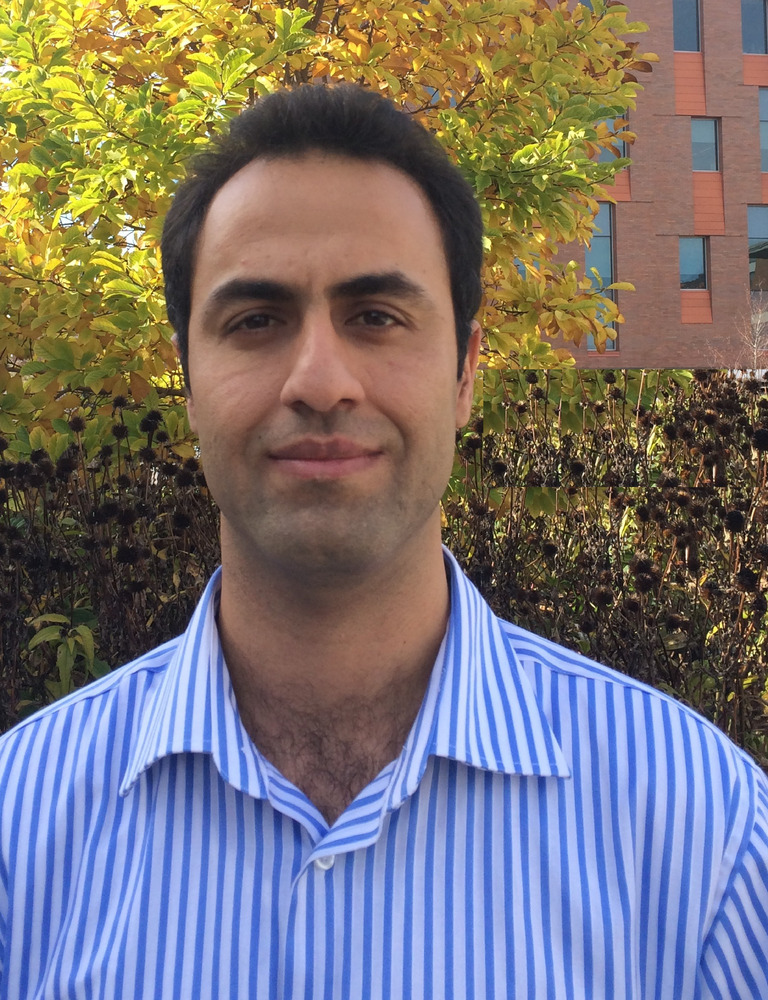}}}]{}
Reza Azarderakhsh received the Ph.D. degree in electrical and computer
engineering from Western University in 2011. He was a recipient of
the NSERC Post-Doctoral Research Fellowship working in the Center
for Applied Cryptographic Research and the Department of Combinatorics
and Optimization, University of Waterloo. Currently, he is a Professor
at Florida Atlantic University. He was the Guest Editor for the \emph{IEEE
Transactions on Dependable and Secure Computing} for the special issue
of Emerging Embedded and Cyber Physical System Security Challenges
and Innovations (2016 and 2017). He was also the Guest Editor for
the \emph{IEEE/ACM Transactions on Computational Biology and Bioinformatics}
for special issue on security. He is serving as an Associate Editor
of \emph{IEEE Transactions on Circuits and Systems (TCAS-I)}.
\end{IEEEbiography}

\end{document}